\documentclass{ws-mpla}
\usepackage[super]{cite}
\usepackage{graphicx}
\begin{document}

\markboth{A.~Oliveros, Enzo L.~Solis, Mario A.~Acero}
{Late cosmic acceleration in a vector--Gauss-Bonnet gravity model}

\catchline{}{}{}{}{}

\title{LATE COSMIC ACCELERATION IN A VECTOR--GAUSS-BONNET GRAVITY MODEL}

\author{\footnotesize A.~OLIVEROS$^{*}$, ENZO L.~SOLIS$^{\dagger}$, MARIO A.~ACERO$^{\ddagger}$}

\address{Programa de F\'isica, Universidad del Atl\'antico, \\ Km 7 antigua via a Puerto Colombia, Barranquilla, Colombia\\
$^{*}$alexanderoliveros@mail.uniatlantico.edu.co\\
$^{\dagger}$enzosolis@mail.uniatlantico.edu.co\\
$^{\ddagger}$marioacero@mail.uniatlantico.edu.co\\}

\maketitle

\pub{Received (Day Month Year)}{Revised (Day Month Year)}

\begin{abstract}
In this work we study a general vector-tensor model of dark energy with a Gauss-Bonnet term coupled to a vector field and without explicit potential terms. Considering a spatially flat FRW type universe and a vector field without spatial components, the cosmological evolution is analysed from the field equations of this  model, considering two sets of parameters. In this context, we have shown that it is possible to obtain an accelerated expansion phase of the universe, since the equation state parameter $w$ satisfies the restriction $-1<w<-1/3$ (for  suitable values of model parameters). Further, analytical expressions for the Hubble parameter $H$, equation state  parameter $w$ and the invariant scalar $\phi$ are obtained. We also find that the square of the speed of sound is negative for all values of redshift, therefore, the model presented here shows a sign of instability under small perturbations. We finally perform an analysis using $H(z)$ observational data and we find that for the free parameter $\xi$ in the interval $(-23.9, -3.46)\times 10^{-5}$, at $99.73\%$ C.L. (and fixing $\eta = -1$ and $\omega = 1/4$), the model has a good fit to the data.

\end{abstract}

\ccode{PACS Nos.: 95.36.+x, 98.80.-k, 04.50.Kd}

\section{Introduction}
\label{sec_intro}

The existence of dark energy (DE) was postulated
by two independent research groups in 1998 (High-z Supernova Search Team and Supernova Cosmology Project) to explain the regime of accelerated expansion of the current universe and became one of the greatest unsolved mysteries in physics; as a matter of fact, its origin from the fundamental point of view is still unknown
\cite{riess, riess2, riess3, riess4, riess5}. This phenomenon has been  supported also by the observed anisotropies in the CMB spectrum \cite{spergel, spergel2} and the analysis of the large scale structures in the universe \cite{tegmark, tegmark2, tegmark3}.  Many  phenomenological and theoretical models have been considered in the last years to resolve the dark energy problem \cite{copeland1}.

The natural candidate for dark energy is the cosmological constant \cite{weinberg,padmanabhan}, conventionally associated
with the energy of the vacuum with constant energy density and pressure, and an equation of state $w =-1$. Other proposals incorporate
dynamical models of DE with scalar fields, which include quintessence, K-essence, tachyon, phantom, ghost condensates and dilaton
\cite{copeland1,piazza}. An alternative to dark energy is related to modified theories of gravity, $f(R)$ \cite{capozziello,nojiri}, or in its more general form $f(R,\mathcal{G})$ \cite{laurentis1,laurentis2}, in which dark energy emerges from the modification of geometry. Besides of scalar fields models of DE, in recent years some authors have considered another alternative about the nature of dark energy, suggesting the possibility that it could be described by a vector field \cite{kiselev,wei,mota,maroto1,maroto2,harko}. In this sense, we propose a vector-tensor model of dark energy with a Gauss-Bonnet (GB) term coupled to vector field and without explicit potential terms. With this proposal, we analyse a possible accelerated expansion phase of the universe in late times. In general, the GB term in four dimensions is known as a topological term, so the field equations are not influenced by it. However, when one considers the GB term coupled to the matter field, the field equations are modified. Further, the GB invariant has an important role in models with extra dimensions (see Ref.~\refcite{farakos} and references therein) and  string theory. In string theory, the GB term arises when the tree-level effective action of the heterotic string is analysed \cite{curtis,gross}. The GB term has been used for several authors in the context of inflation and dark energy  models. A model with vector field coupled to GB invariant in a type Bianchi I universe was proposed in Ref.~\refcite{mota}. Recently, in Ref.~\refcite{valenzuela} the authors have
considered an inflation model with a cosmological vector field non-minimally coupled to gravity through the GB invariant. 
In the literature, usually the GB invariant has been used in models with scalar fields
\cite{odintsov,calcagni,carter,sami,neupane,mota2,odintsov2,bamba,odintsov3,odintsov4,granda}, and in these papers it is shown that the GB coupling term may be relevant for the explanation of accelerated expansion of the universe at late times.

This paper it is organized as follows: in section \ref{sec_model} we introduce the vector-tensor model with a GB term coupled to the vector field and the 
corresponding field equations are obtained. In section \ref{sec_cosmoEvol}, we consider a flat FRW type universe and a vector field without spatial components, and from these considerations we analyse the cosmological evolution of this model. Section \ref{sec_fit} is devoted to a statistical analysis of the model, by performing a fit of the single model parameter $\xi$, to the observational data for $H(z)$. Finally, some conclusions are exposed in section \ref{sec_concs}.

\section{The model}
\label{sec_model}
The most general action for the vector-Gauss-Bonnet dark energy model (without mater contributions) can be written as
\begin{equation}\label{eq1}
\begin{aligned}
S = -\int d^4x \sqrt{-g}\Bigl[\frac{R}{2} 
    +\frac{1}{4}F_{\mu \nu} F^{\mu \nu}  
    +&\,\eta R_{\mu \nu}A^{\mu}A^{\nu} 
    +\omega R A_{\mu}A^{\mu} \\
    +&\,\xi \mathcal{G} A_{\mu}A^{\mu}+V(A^2)\Bigr],
\end{aligned}
\end{equation}
where $F_{\mu \nu}=\partial_{\mu}A_{\nu}-\partial_{\nu}A_{\mu}$, $R_{\mu\nu}$ and $R$ are the Ricci tensor and the Ricci scalar, respectively, $V(A^2)$ is a potential term, $\mathcal{G}=R^2-4R_{\mu\nu}R^{\mu\nu}+R_{\mu\nu\alpha\beta}R^{\mu\nu\alpha\beta}$ is the topological GB invariant, and free coupling constants ($\eta$, $\omega$, $\xi$) for each non-minimal coupling term are included. $\omega$ and $\eta$ are dimensionless, but the dimensions of $\xi$ are $M^{-2}$.

Different specific forms of this action have been studied, for instance, in Ref. \refcite{harko}, where dark energy is proposed to be represented by a massive vector field non-minimally coupled to gravity, but the Gauss-Bonnet invariant is not considered ($\xi = 0$ and 
$V(A^2)=\frac{1}{2}\mu^2_{\Lambda}A_{\mu}A^{\mu}$ in Eq.~(\ref{eq1})); also, the authors of Ref. \refcite{sadeghi} considered a modified Gauss-Bonnet gravity model ($\eta = \omega = 0$, $V(A^2)=0$ in Eq.~(\ref{eq1})), in which inflation and late-time acceleration of the universe is realised.

\noindent The variation of the action (\ref{eq1}) with respect to the metric tensor $g_{\mu\nu}$ gives the field equations
\begin{equation}\label{eq7}
R_{\mu \nu} - \frac{1}{2}g_{\mu \nu}R = \kappa^2 T_{\mu \nu}^{DE},
\end{equation}
where $T^{DE}_{\mu \nu}$ is the energy momentum tensor for the vector field and has the following form:
\begin{equation}\label{eq6}
T^{DE}_{\mu \nu} = T^{F^2}_{\mu \nu}+T^{RAA}_{\mu \nu}+T^{\mathcal{G}}_{\mu \nu}+T^{R \phi}_{\mu \nu}+T^{V(\phi)}.
\end{equation}
Each term is given by the following expressions (Eqs.~(\ref{eq2})-(\ref{eq5a})):
\begin{equation}\label{eq2}
T^{F^2}_{\mu \nu} = F_{\mu \beta} F^{\,\,\,\, \beta}_{\nu} -\frac{1}{4} g_{\mu \nu} F_{\alpha \beta} F^{\alpha \beta},
\end{equation}

\begin{equation}\label{eq3}
\begin{aligned}
T^{RAA}_{\mu \nu} = \eta \Bigl( g_{\mu \nu} \Bigl[R_{\alpha \beta} A^{\alpha} A^{\beta} &\,- \nabla_{\alpha} \nabla_{\beta} (A^{\alpha} A^{\beta}) \Bigr]  - \square(A_{\mu}A_{\nu}) \\ &\,+ 2\nabla_{\beta} \nabla_{(\mu} (A_{\nu)} A^{\beta}) - 4 R_{\beta (\mu } A_{\nu)} A^{\beta} \Bigr),
\end{aligned}
\end{equation}

\begin{equation}\label{eq4}
\begin{aligned}
T^{\mathcal{G}}_{\mu \nu} = \xi \Bigl( 8\Bigl[ R^{\,\,\, \alpha \beta}_{\mu \,\,\,\,\, \nu} \nabla_{\alpha} \nabla_{\beta} (\phi) &\,+ R_{\mu \nu} \square \phi - 2\nabla_{\beta} \nabla_{(\mu} (\phi)R_{\nu)}^{\,\,\,\beta} + \frac{1}{2}R \nabla_{\mu} \nabla_{\nu} \phi \Bigr] \\
&\,+  4\Bigl[ 2 R^{\alpha \beta} \nabla_{\alpha} \nabla_{\beta} (\phi) -R \square(\phi) \Bigr] g_{\mu \nu} -2\mathcal{G} A_{\mu} A_{\nu} \Bigl),
\end{aligned}
\end{equation}

\begin{equation}\label{eq5}
\begin{aligned}
T^{R \phi}_{\mu \nu} =  -2 \omega\Bigl[ R A_{\mu}A_{\nu} + \Bigl(R_{\mu \nu} - \frac{1}{2} g_{\mu \nu} R \Bigr) \phi + g_{\mu \nu} \square (\phi) - \nabla_{(\mu} \nabla_{\nu)} (\phi)  \Bigl],
\end{aligned}
\end{equation}

\noindent and 
\begin{equation}\label{eq5a}
T^{V(\phi)}_{\mu \nu} = -2 \frac{d V(A_{\kappa} A^{\kappa})}{d A_{\kappa} A^{\kappa}} A_{\mu} A_{\nu} + g_{\mu \nu} V(A_{\kappa} A^{\kappa})
\end{equation}
where $\phi=A_{\mu}A^{\mu}$, is an invariant scalar. For more details about the variation of the Gauss-Bonnet term, see \cite{odintsov}.

On the other hand, the variation of the action with respect to $A_{\mu}$, gives the equation of  motion 
\begin{equation}\label{eq8}
-\nabla_{\mu} F^{\mu \nu}+ 2 \eta R^{\nu}_{\,\,\mu} A^{\mu} + 2 \omega R A^{\nu}- 2 \xi \mathcal{G} A^{\nu}+ 2 \frac{dV(A_\kappa A^\kappa)}{dA_\kappa A^\kappa} =  0 .
\end{equation}

\section{Cosmological evolution}
\label{sec_cosmoEvol}
\noindent To analyse the cosmological evolution of the universe generated by the previous model and determine if it is possible to obtain a regime of accelerated expansion of the universe at late times, we regard the flat Robertson-Walker metric for a homogeneous and isotropic universe, given by
\begin{equation}\label{eq12}
ds^2=dt^2-a(t)^2\sum_{i=1}^{3}{(dx_i)^2},
\end{equation}
where $a(t)$ is the scale factor. We consider that the vector field has only the temporal component, i.~e.~$A^{\mu}=(A^0(t),0,0,0)$, which is in agreement with the observed isotropy and homogeneity of the universe.\\

\noindent Using the FRW metric in Eq.~(\ref{eq8}) and taking $V(A^2)=0$, we obtain 
\begin{equation}\label{eq8a}
(\eta+2\omega)\dot{H}A_0+(\eta+4\omega)H^2A_0+8\xi H^2(\dot{H}+H^2)A_0=0.
\end{equation}
Here one can examine a number of interesting possibilities in order to simplify the problem under study; for instance, with $\eta+4\omega=0$, then $\eta=-4\omega$, so that if $\omega=1/4$ then $\eta=-1$. In the present work, we have decided to study this case to analyse the cosmological evolution of such a model. Further details of this choice are discussed bellow.

From Eq.~(\ref{eq7}) and using the previous considerations, the Friedmann equations take the following form:
\begin{equation}\label{eq9}
3H^2=\rho_A,
\end{equation} 
and
\begin{equation}\label{eqfrie}
\begin{aligned}
2\dot{H}+3H^2=-p_A.
\end{aligned}
\end{equation}
Here we have used $\kappa^2=1$. $\rho_A$ and $p_A$ are the dark energy and pressure density respectively of the vector field and are given by (using $\omega = 1/4$, $\eta = -1$)
\begin{equation}\label{eqDE}
\begin{aligned}
\rho_A = 3A_{0}^2 \dot{H}-\frac{9}{2} A_{0}^2 H^2 - 3 A_{0} \dot{A_{0}} H + 48 \xi [A_{0} \dot{A_{0}} H^3 - A_{0}^2 H^2(\dot{H}+H^2 )],
\end{aligned}
\end{equation}
and
\begin{equation}\label{eq10}
\begin{aligned}
p_A = \dot{A}_{0}^{2}+A_0\ddot{A}_0 + 6A_0\dot{A}_0H 
    + &A_{0}^{2}\Bigl(\frac{9}{2}H^2+3\dot{H}\Bigr) \\
    - &16\xi[H^2(\dot{A}_{0}^{2}+A_0\ddot{A}_0)
    + 2A_0\dot{A}_0H(\dot{H}+H^2)].
\end{aligned}
\end{equation}

\noindent On the other hand, with the selected values for the coupling constants, the equation of motion (\ref{eq8a}) becomes
\begin{equation}\label{eq11}
-\frac{1}{2}\dot{H} A_{0} + 8 H^2 (\dot{H}+H^2)A_{0} \xi = 0,
\end{equation}
from which it is possible to obtain an analytical expression for the Hubble parameter, since the component $A_0$ of the vector
field is easily decoupled (a similar result was obtained in Ref.~\refcite{harko}). To facilitate the procedure, we carry out the change of variable $x=\ln{a}$ 
($x$ is  known as the e-folding variable). Then, Eq.~(\ref{eq11}) can be rewritten as
\begin{equation}\label{eq13}
-\frac{1}{4} \frac{dH^2}{dx}  + 8 H^2 \left[\frac{1}{2} \frac{dH^2}{dx} +H^2\right] \xi = 0,  
\end{equation}
and the solution for this differential equation is
\begin{equation}\label{eq14}
H^2(x) = - \frac{1}{16 \xi} \left[W\left(- \frac{\exp\left[2 x - \frac{A}{16 \xi}\right]}{16 \xi}\right)\right]^{-1}, 
\end{equation}
where $A$ is a integration constant and $W$ is the ``Lambert W-Function'' or ``ProductLog'' function. Using the initial condition, $H(0)=H_0$ in (\ref{eq14}), the integration constant is
\begin{equation}\label{eq15}
A=\frac{1}{H_0^2}+16\xi\ln{H_0^2}.
\end{equation}

Before continuing with the analysis of this model, we notice that by looking at Eq.~(\ref{eq8a}), it is easy to see that another possible choice is $\eta+2\omega=0$. In this case, $\eta=-2\omega$, so that if $\omega=1/2$ one gets $\eta=-1$. With these values for the coupling constants, the first term in Eq.~(\ref{eq8a}) is neglected, and the Einstein's tensor $G_{\mu\nu}=R_{\mu\nu}-\frac{1}{2}g_{\mu\nu} R$, arises in the action Eq.~(\ref{eq1}); also, from  Eq.~(\ref{eq8a}), the resulting differential equation has the solution
\begin{equation}\label{eq8b}
H^2(x)=\frac{e^{-2x}(1+8\xi H_0^2-e^{2x})}{8\xi}.
\end{equation}
We stress that we checked the fact that the two exposed choices do produce very similar results, even though the solution shown in Eq.~(\ref{eq8b}) is different compared to the Eq.~(\ref{eq14}).

Now, we consider the behavior of the equation of state parameter $w$ in order to determine if it is possible to obtain an accelerated expansion regime for the universe at late times. The equation of state parameter $w$ is given by
\begin{equation}\label{eqw1}
w=-1-\frac{1}{3H^2(x)}\frac{d}{dx}H^2(x),
\end{equation}
and using (\ref{eq14}) and (\ref{eq15}), we obtain
\begin{equation}\label{eqw2}
w(x)=-1+\frac{2}{3}\left[1+ W\left(-\frac{\exp\left[2 x - \frac{1}{16 H_{0}^2 \xi}\right]}{16 H_{0}^2 \xi}\right)\right]^{-1}.
\end{equation}

\begin{figure}[h]
\centerline{\includegraphics[scale=0.42,angle=-90]{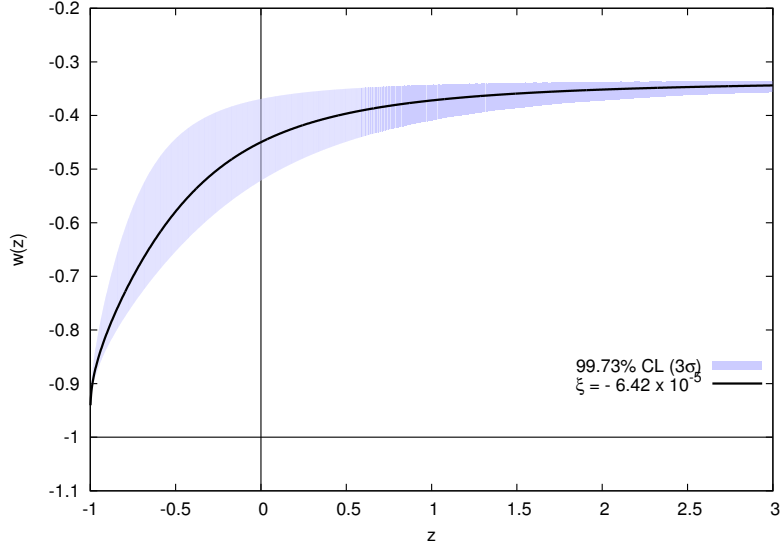}}
\caption{Evolution of $w$ as a function of $z$, using the value of the free parameter which gives a best fit to the data. A $3\sigma$ (shadowed) region for allowed values of $\xi$ is shown. \protect\label{fig1}}
\end{figure}

\noindent Fig.~\ref{fig1} shows the behavior of $w$ as a function of the redshift $z$. The continuous line corresponds to the evolution of $w$ for $\xi_{BF}$, while the shaded region shows the result of $w$ for different possible values of $\xi$ within a $3\sigma$ ($99.73\%$ C.L.) interval (see Sec.~\ref{sec_fit} for the details). We have considered here $H_0=67.8\,\rm{km}\,\rm{s}^{-1}\,\rm{Mpc}^{-1}$ \cite{planck}. Also, the substitution $x=-\ln{(1+z)}$ has been realized.

\noindent We can see in the Fig.~\ref{fig1} that it is possible to obtain a regime of accelerated expansion of the universe at late times for a wide range of values of the parameter $\xi$ (up to a $99.73\%$ CL), since the equation state parameter satisfy the restriction $-1<w<-1/3$. This model behaves as a quintessence scalar field model and a phantom phase is not observed ($w$ does not cross the $w = -1$ barrier). Furthermore, the function $w(z)$ satisfies the following limits 
\begin{equation}\label{eqwl}
\lim_{z\rightarrow\infty}w(z)=-\frac{1}{3},\ \ \ \ \ \lim_{\xi\rightarrow 0}w(z)=-1.
\end{equation}

\noindent The second limit corresponds to a de Sitter solution ($H$ is constant). In this case, the associated cosmological constant arises from the vector field. Finally, 
\begin{equation}\label{eqwl2}
\lim_{z\rightarrow -1}w(z)=-1.
\end{equation}
In this last limit, the value of coupling constant $\xi$ is arbitrary and in this stage, the universe is dominated by the cosmological constant (de Sitter phase).

\noindent From Eq.~(\ref{eq9}) and using (\ref{eqDE}) and (\ref{eq15}), we can obtain an expression for the invariant scalar $\phi=A_0A^0$, obtaining
\begin{equation}\label{eq16}
\phi(x)=-\frac{2}{3}+B\,\exp\left[{-\frac{3}{2}W\left(-\frac{\exp\left[2 x - \frac{1}{16 H_{0}^2 \xi}\right]}{16 H_{0}^2 \xi}\right)}\right],
\end{equation}
where $B$ is an integration constant, which is given by the initial condition $\phi(0)=\phi_0$. The value of initial condition must be consistent with the restriction $\phi>0$.\\ 

\begin{figure}[ph]
\centerline{\includegraphics[scale=0.42,angle=-90]{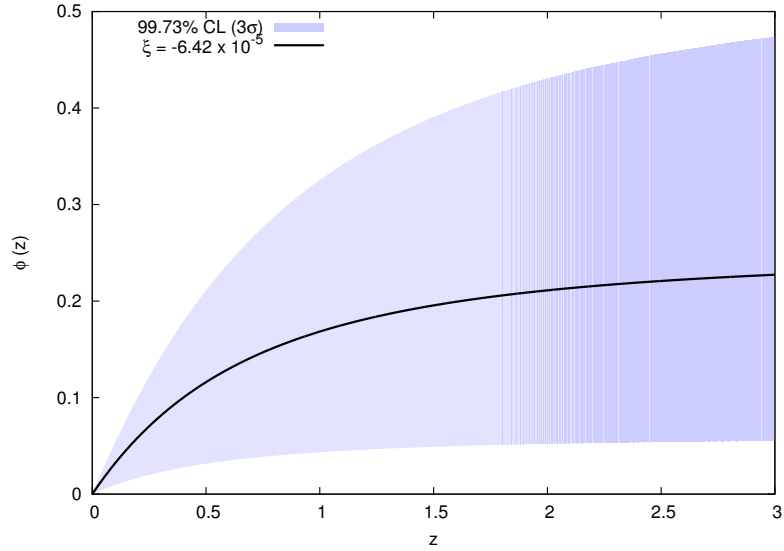}}
\caption{The invariant scalar $\phi$ versus the redshift $z$. See Fig.~\ref{fig1} for the explanation of the shadowed area.\label{fig2}}
\end{figure}

\noindent Choosing initial condition, $\phi(0)=0$, from (\ref{eq16}) we obtain 
\begin{equation}\label{eq17B}
B=\frac{2}{3}\exp\left[\frac{3}{2}W\left(-\frac{\exp\left[ - \frac{1}{16 H_{0}^2 \xi}\right]}{16 H_{0}^2 \xi}\right)\right].
\end{equation}
Then, with (\ref{eq16}) we plot the scalar invariant $\phi$ vs $z$ in Fig.~\ref{fig2}. We again use the $\xi_{BF}$ value for the free parameter (explanation provided in Sec.~\ref{sec_fit}) and the corresponding $3\sigma$ allowed interval.
We see that the function $\phi$ is always positive for all redshift  values.\\

\begin{figure}[ph]
\centerline{\includegraphics[scale=0.42,angle=-90]{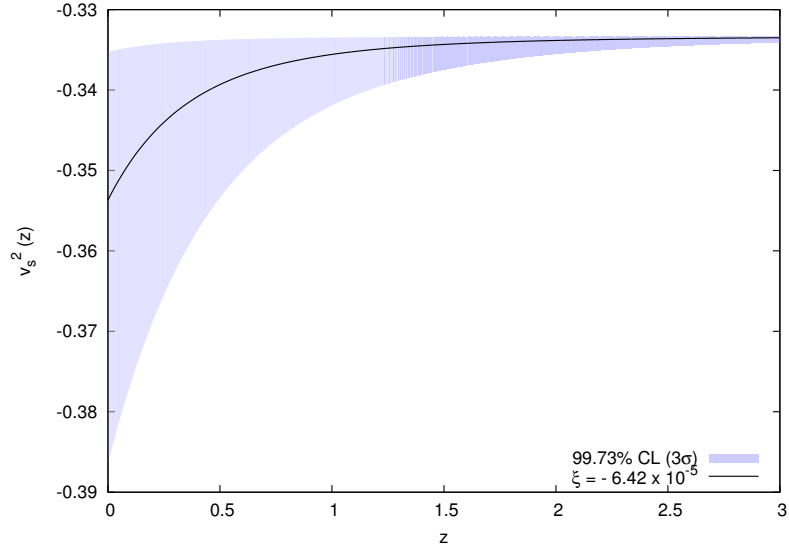}}
\caption{Plot of the square of the speed of sound $v_s^2$ as a function of $z$. See Fig.~\ref{fig1} for the explanation of the shadowed area. \label{fig3}}
\end{figure}

In order to examine the stability of the model, we study the square of the speed of sound ($v_s^2$) as a function of the
redshift. $v_s^2$ is given by \cite{peebles}
\begin{equation}\label{eq17}
v_s^2=\frac{\dot{p}}{\dot{\rho}}=\frac{p'}{\rho'},
\end{equation}
where the prime means differentiation with respect to $x$. From of Eqs.~(\ref{eqDE}), (\ref{eq10}), (\ref{eq14}) and (\ref{eq16}), Eq.~(\ref{eq17}) takes the explicit form
\begin{equation}\label{eq18}
\begin{aligned}
v_s^2(x)=&-1-\frac{2}{3}\left[1+ W\left(-\frac{\exp\left[2 x - \frac{1}{16 H_{0}^2 \xi}\right]}{16 H_{0}^2 \xi}\right)\right]^{-2}\\
&+\frac{4}{3}\left[1+ W\left(-\frac{\exp\left[2 x - \frac{1}{16 H_{0}^2 \xi}\right]}{16 H_{0}^2 \xi}\right)\right]^{-1},
\end{aligned}
\end{equation}
and its evolution is depicted in Fig.~\ref{fig3}, including the $3\sigma$ shaded band for the free parameter $\xi$, as explained before. 
It is noted that $v_s^2$ is  negative for all values of z, showing that the model presented here shows a sign of instability under small perturbations. This behavior does not depend of the initial condition $\phi(0)=\phi_0$, which is clear from Eqs.~(\ref{eqDE}), (\ref{eq10}) and (\ref{eq17}) (one must take into account $\phi=A_0^2$, $\dot{\phi}/2 = A_0\dot{A}_0$ and $\ddot{\phi}/2 = \dot{A}_0^2+A_0\ddot{A}_0$). However, it is important to note that the positivity of $v_s^2$ is a necessary condition but is not enough to conclude that the model is stable \cite{sheykhi}. 

\section{Observational constraints using $H(z)$ data}
\label{sec_fit}
A simple but robust procedure to study models containing free parameters, is to perform a least-squares analysis (see Section 38 of Ref.~\refcite{pdg}), comparing theoretical predictions from the model for a specific measurable quantity, with experimental or observational data. In this case, one can carry out such an analysis using the observational data of $H(z)$ and the obtained result for the Hubble parameter given by Eq.~(\ref{eq14}) (or Eq.~(\ref{eq8b}) -see bellow-), which depends explicitly on the only free parameter of the proposed model.

Here we consider data of the Hubble parameter $H(z)$ for different values of the redshift, taken from Table I of Ref.~\refcite{Ding}, and consider the simple $\chi^2$-function
\begin{equation}
\chi^2 = \sum_{i=1}^{N}\frac{\left(H_{\rm{the}}^i(z;\xi) - H_{\rm{obs}}^i(z)\right)^2}{(\sigma_{\rm{obs}}^i(z))^2},
\end{equation}
where the sum runs up to $N=29$, the total number of observed values; $H_{\rm{obs}}(z)$ and $\sigma_{\rm{obs}}(z)$ are the observed $H(z)$ and its corresponding uncertainty, respectively (as in \cite{Ding}); and $H_{\rm{the}}(z;\xi)$ is the theoretical Hubble parameter, which depends on the redshift, $z$, as well as on the free parameter $\xi$, as expressed by (\ref{eq14}).

\begin{figure}[hp]
\centerline{\includegraphics[scale=0.42,angle=-90]{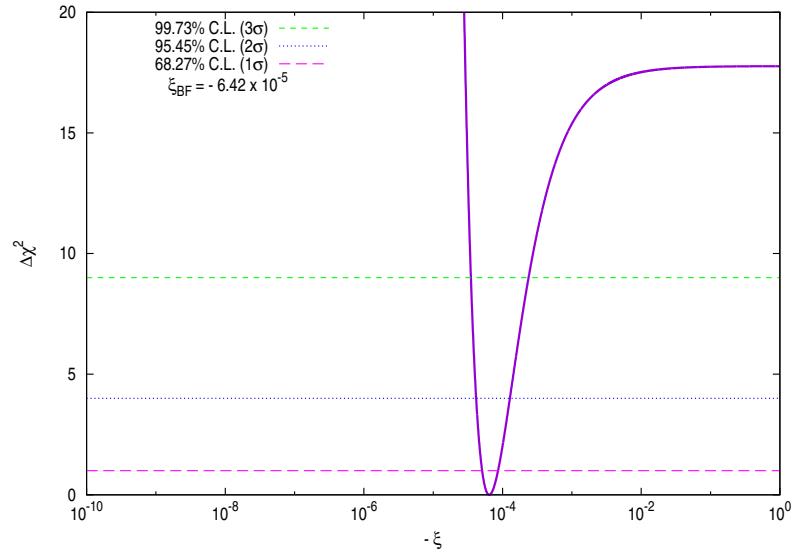}}
\caption{Allowed values of the $\xi$ parameter obtained from the least-squares analysis of the $H(z)$ data. The best fit value corresponds to $\Delta \chi^2 = 0$.}
\label{fig_chi2}
\end{figure}

The main result of this analysis is presented in the form of the curve show in Fig.~\ref{fig_chi2}, where the quantity $\Delta\chi^2 \equiv \chi^2 - \chi^2_{\rm{min}}$ is plotted for an interesting interval of values of the free parameter $\xi$. We found that the fit produces 
$\chi^2_{\rm{min}} = 34.85$ (for 28 d.o.f) when 
\begin{equation}\label{eq_BF}
\xi_{\rm{BF}} = -6.42 \times 10^{-5}. 
\end{equation}

In the figure, and according to \cite{pdg}, we also show the allowed intervals for the $\xi$ parameter up to $1\sigma$ (pink long--dashed line), $2\sigma$ (blue dotted line) and $3\sigma$ (green short--dashed line). The corresponding intervals can be seen in Table \ref{tab_xi}.

\begin{table}[h]
\tbl{$1,2,3 \, \sigma$ intervals for the $\xi$ parameter obtained by the data fit of the model to $H(z)$ data.}
{\begin{tabular}{@{}cc@{}} \toprule
   C.L.   & $\xi$ \\ \colrule
$68.27\% \, (1\sigma)$ & $(-8.87, -5.02)\times 10^{-5}$ \\
$95.45\% \, (2\sigma)$ & $(-13.2, -4.11)\times 10^{-5}$ \\
$99.73\% \, (3\sigma)$ & $(-23.9, -3.46)\times 10^{-5}$ \\ \botrule
\end{tabular}
\label{tab_xi}}
\end{table}%

Notice that our analysis excludes the case for which the GB term is not coupled to the vector field, $\xi = 0$, at more than $99.73\%$ C.L. Also, despite of the apparent narrowness of the curve in fig.~\ref{fig_chi2} (note its log-scale of the horizontal axis), we obtained wide ranges for the values of the free parameter. In order to get even stronger constraints, one would need to have observational information with smaller uncertainties, or to use data of some other observables. For instance, one could use observational bounds on $H_0$, as this parameter also appears explicitly in the theoretical prediction for $H(z)$, equations (\ref{eq14}) and (\ref{eq15}). Otherwise, PPN solar system constraints could be implemented.

\begin{figure}[hp]
\centerline{\includegraphics[scale=0.42,angle=-90]{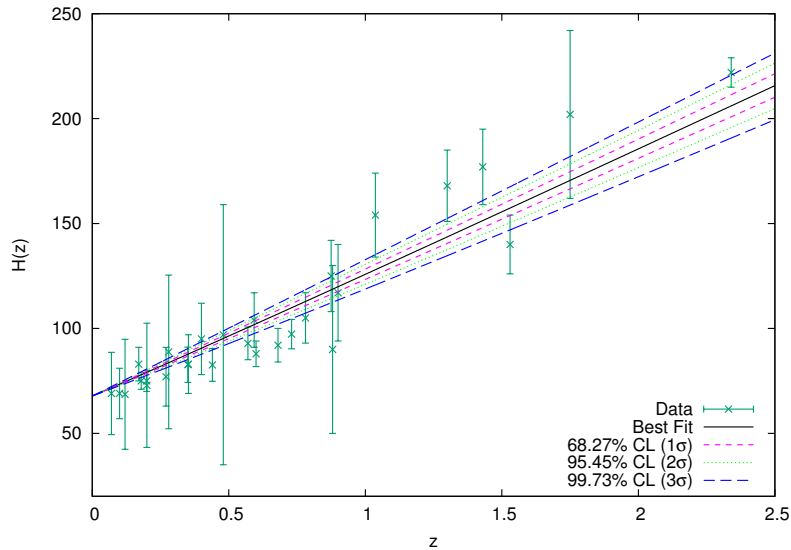}}
\caption{$H(z)$ as given by Eq.~(\ref{eq14}) for $\xi_{BF}$ obtained by the data fit. Observational data are also plotted to show the goodness of the fit, including $1\sigma$, $2\sigma$ and $3\sigma$ C.L. bands for the fitted parameter (i.e., using values of $\xi$ as in table \ref{tab_xi} in Eq.~(\ref{eq14}).)}
\label{fig_fit}
\end{figure}

In order to have a taste of the goodness of our fit, we compare the model prediction for $H(z)$ using the best fit value obtained for the free parameter, $\xi_{\rm{BF}}$, Eq.~(\ref{eq_BF}), with the data points from \cite{Ding}. The resulting plot is shown in Fig.~\ref{fig_fit}. 

In the figure, data are plotted with their uncertainties and in addition to the Best Fit curve (continuous line), the $3\sigma$ (inside the blue long--dashed lines), $2\sigma$ (inside the green dotted lines) and $1\sigma$ (inside the pink short--dashed lines) bands are also depicted. Fig.~\ref{fig_fit} clearly shows that data points (plus uncertainties) are very well covered by the $3-$ and even the $2\sigma$ bands, demonstrating that our model exhibits a good fit to the data. Once again, stronger bounds on observational data would certainly allow one to strengthen also the limits on the model.

\begin{figure}[h]
\centerline{\includegraphics[scale=0.42,angle=-90]{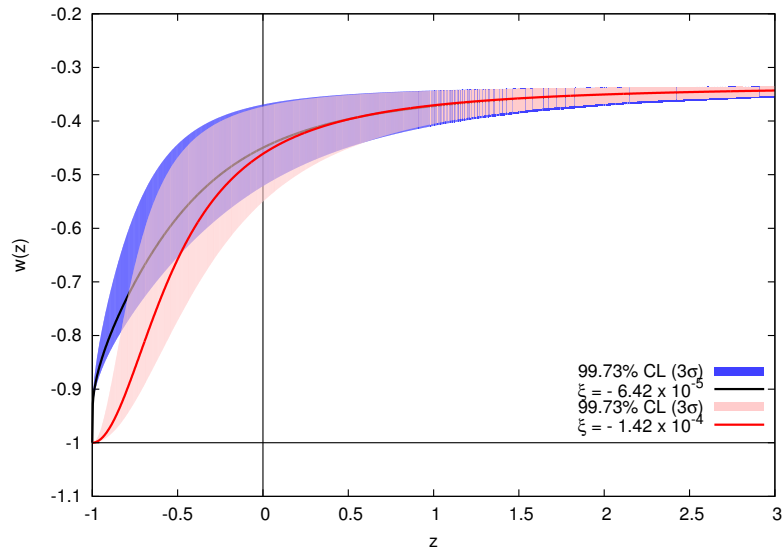}}
\caption{Comparison of the evolution of $w$ as a function of $z$ for the two possible solutions discussed on Section \ref{sec_cosmoEvol}. The blue (darker) zone corresponds to ($\eta=-1, \omega=1/4$), as in Fig.~\ref{fig1}, while the red (lighter) region corresponds to ($\eta=-1, \omega=1/2$).}
\label{fig_compare}
\end{figure}

Finally, as anticipated above, we have also performed the same kind of analysis for the other possible values of the parameters in Eq.~(\ref{eq8a}), i.e. $\eta=-1$ and $\omega=1/2$. The fit to the observational data is very similar, giving a best fit value for the free parameter $\xi = -1.42 \times 10^{-4}$ (with $\chi^2_{\rm{min}}=31.33 / 28 \, \rm{ d.o.f}$). Although this value is slightly larger than (\ref{eq_BF}), the evolution of $H(x)$ given by Eq.~(\ref{eq8b}) is almost equal to the one given by Eq~(\ref{eq14}). As a result, and in order to show this similarity, we compare the evolution of the equation of state, $w$, as a function of the redshift. The result is exposed on Fig.~\ref{fig_compare}.
It is clear from the figure that the behaviour of $w(z)$ is basically the same for the two sets of parameters ($\eta, \omega$), from early times until present. In both of the cases, it remains valid that $w < -1/3$, as expected. On the other hand, even though there is a visible difference in the evolution of the two plot for $z < 0$, both of them go to $-1$, as expected, as well.

With this in hands, anyone could ask why we decided to work and exhibit one set of parameters over the other. In deed, the main reason is the fact that the case of ($\eta=-1, \omega=1/2$) does not give an analytical solution for the invariant scalar $\phi$, so making the case of ($\eta=-1, \omega=1/4$) more interesting and easier to analyse.

\section{Conclusions}
\label{sec_concs}
\noindent The GB invariant coupled to the scalar fields has been widely used in the literature and it is shown that the GB coupling term may be relevant for the explanation of accelerated expansion of the universe at late times; it has also been considered in inflationary scenarios.
Furthermore, the GB invariant has an important role in models with extra dimensions (braneworlds) and in string theory. 

In this paper we have shown that it is possible to obtain an accelerated expansion phase of the universe for two different sets of values of the parameters $\eta$ and $\omega$, from a general vector-tensor model given by Eq.~(\ref{eq1}), where a vector field coupled to GB term has been included. This conclusion is supported by the equation state parameter behavior shown in Fig.~\ref{fig1} (and Fig.~\ref{fig_compare}), as well as Eq.~(\ref{eqw2}), satisfying the restriction $-1<w<-1/3$. Since $w$ does not cross $w=-1$ barrier, then, a phantom phase isn't observed. Moreover, when $z\rightarrow \infty$ the equation state parameter $w\rightarrow -1/3$ and when $\xi\rightarrow 0$, we have that $w=-1$. This latter case corresponds to a de Sitter solution ($H$ is constant). In this case, the associated cosmological constant arises from the vector field. When $z\rightarrow -1$ (future) then $w\rightarrow -1$, for any value of coupling constant $\xi$ and the universe in this case is dominated by the cosmological constant (see eqs.~(\ref{eqwl}) and (\ref{eqwl2})). We have also obtained an analytical expression for the invariant scalar $\phi=A_0A^0$ given by Eq.~(\ref{eq16}) and the initial condition $\phi(0)=0$ was considered, which agrees with the restriction $\phi>0$ (see Fig.~\ref{fig2}).

In order to examine the stability of the model, the square of the speed of sound ($v_s^2$) as a function of the redshift was analysed (see (\ref{eq18})) and we found that the square of the speed of sound is  negative for all values of redshift; under this criterion, the model presented here shows a sign of instability under small perturbations (see Fig.~\ref{fig3}). This behavior does not depend of the initial condition $\phi(0)=\phi_0$. 

Finally, by performing a least-squares analysis with observational data of the Hubble parameter $H(z)$, we found that our model exhibits a good match to the data with a best fit given by $\xi_{\rm{BF}} = -6.42 \times 10^{-5}$ for $\eta = -1$ and $\omega = 1/4$ (and $\xi_{\rm{BF}} = -1.42 \times 10^{-4}$ for $\eta = -1$ and $\omega = 1/2$). As explained in Sec.~\ref{sec_fit}, the analysis provides us with allowed regions for the free parameter $\xi$ (see Fig.~\ref{fig_chi2} and Table \ref{tab_xi}), which are used to compute the important quantities along this article ($w(z)$, $\phi(z)$ and $v_s^2(z)$), as well as to perform a comparison with the data (Fig.~\ref{fig_fit}). Accordingly, we can say that the vector-tensor model of dark energy considered in this work is a phenomenologically viable model. We stress out that, in a future work, the free parameters of the model (the coupling constant and initial value of the vector field) could be  estimated by using additional restrictions, as for example, the PPN solar system constraints.



\end{document}